\documentclass[pra,twocolumn]{revtex4}
\usepackage{amsmath}
\usepackage{amssymb}
\usepackage{mathtools}
\usepackage{graphicx}

\renewcommand{\vec}[1]{\mathbf{#1}}
\newcommand{\mat}[1]{{\underbracket[0.5pt][0pt]{\mathbf{#1}}}}
\newcommand{\dif}{\mathrm{d}}
\DeclareMathOperator{\Tr}{Tr}
\DeclareMathOperator*{\sump}{{\sum\nolimits^{\prime}}}

\makeatletter
\gdef\@ptsize{0}
\let\@currsize\normalsize
\makeatother

\newlength{\fullwidth}
\newlength{\halfwidth}
\begin{document}

\title{Calculating Casimir interactions for Periodic Surface Relief
  Gratings using the C-Method}

\author{Jef Wagner} 
\email{jeffrey.wagner@ucr.edu}

\author{Roya Zandi}
\email{roya.zandi@ucr.edu}
\affiliation{Department of Physics and Astronomy, University of
  California Riverside}


\begin{abstract}
  We develop a formalism to calculate the fluctuation-induced
  interactions in periodic systems. The formalism, which combines the
  scattering theory with the C method borrowed from electromagnetic
  gratings studies, is suitable and efficient for the calculation of
  the Casimir forces involving surface relief gratings. We apply the
  developed technique to obtain the energy and lateral force for
  simple 1-D sinusoidal gratings. Using this formalism we derived
  known asymptotic expressions that were previously obtained through
  perturbative approximations.  At close separation, our numerical
  results match those obtained by the proximity force approximation
  and its first correction using the derivative expansion.
\end{abstract}

\maketitle

\section{Introduction}
In a seminal paper in 1948, H.~G.~B.~Casimir \cite{Casimir:1948dh}
found the presence of an attractive interaction between two neutral
perfect mirrors in vacuum. This effect was later generalized to real
materials (finite conductors and dielectrics) by
E.~M.~Lifshitz\cite{Lifshitz:1956zz}, which has some consequences for
micro and nano-scale mechanical devices, often leading to a very
strong attraction between parts called
stiction\cite{Serry:1998a,Buks:2000a}. To this end, a complete
quantitative understanding of the Casimir effect is necessary for the
proper design and analysis of MEMS and NEMS. The goal is to be able to
exert some control over the Casimir forces by manipulating the
material or geometry of the system.

There have been a number of experiments showing that the magnitude of
the Casimir force could be varied by changing the surface geometry of
the interacting objects \cite{Roy:1999zz, Chen:2002zzb, Chen:2002zzc,
  Chui:2009a, Chui:2010a, Bao:2010a, Chan:2008a, Intravaia:2013b,
  Banishev:2013a, Banishev:2014a}. To investigate the impact of
curvature and corrugation, the Casimir forces have been measured
between a sphere and a sinusoidal grating \cite{Roy:1999zz} and
between two corrugated surfaces for both aligned \cite{Chen:2002zzb,
  Chen:2002zzc} and crossed \cite{Banishev:2013a, Banishev:2014a}
corrugations.

Until recently most of the theoretical analysis of the Casimir force
experiments has been done using the proximity force approximation
(PFA), also known as the Derjaguin approximation \cite{Derjaguin:1934}
for which the curved surfaces are assumed to be made up of
infinitesimal finite plates.  Using this approximation, one can
calculate the force (or energy) between curved objects through the
expression for infinite parallel plates. This approximation is only
valid in the limit as the separation is much smaller than the radius
of curvature of the curved surfaces. As the experiments have become
more sensitive, it has become necessary to do the theoretical analysis
outside the range of validity of the PFA. Recently a derivative
expansion (DE) approach has been introduced \cite{Fosco:2011xx,
  Bimonte:2012a, Bimonte:2012b}, which reproduces the PFA and gives
the next order correction, and which has successfully explained a
number of new recent experiments \cite{Banishev:2013a,
  Banishev:2014a}.
The first theoretical calculation of the Casimir force between
geometrically patterned surfaces not using the PFA was reported in
2001 \cite{Emig:2001dx, Emig:2002xz}, which described the normal and
lateral forces between two aligned corrugated surfaces in terms of a
perturbative expansion in the profile height. Since then, this
perturbative approach has been expanded to include real material
properties and unaligned corrugations \cite{Neto:2005zz,
  Rodrigues:2006ku, Rodrigues:2006cf, CaveroPelaez:2008tj,
  CaveroPelaez:2008rt}.  More recently the scattering method has been
used to obtain the Casimir forces in periodic systems
\cite{Lambrecht:2009zz, Davids:2010a, Intravaia:2012a}.  For example
in Ref.~\cite{Davids:2010a}, the scattering method along with rigorous
coupled wave analysis (RCWA), an approach developed for
electromagnetic grating theory, is used to calculate the Casimir
forces in the Corrugated systems.

In this paper, we combine the scattering theory with the C method, an
efficient technique for calculating the Rayleigh coefficients
optimized for surface relief gratings \cite{Chandezon:1980a, Li:1999a,
  Chandezon:2002a, Poyedinchuk:2006a}. We note that for smooth height
profiles (such as a sinusoidal grating) the C method has some
significant advantages over the RCWA. The RCWA assumes that the system
is made up layers with square sides, and many layers are required to
accuratly model smooth surface profiles.  In
Ref.~\cite{vanderAA:2004a} the C method was compared to the RCWA for
sinusoidal gratings, and it was found that 40 layers were required in
the RCWA to match the same accuracy from the C method. With 40 layers,
the calculation employing the RCWA was much slower than that using the
C method. It should be noted that for rectangular gratings the RCWA
would perform much better than the C method. This method is not meant
to replace the RCWA, only complement it by describing a method
appropriate for smoothly varying surfaces.

The structure of the paper is as follows. While Sect.~\ref{sec:scat}
briefly describes the scattering method, and explicitly gives the
basis functions and the translation matrix, Sect.  \ref{sec:Cmeth}
describes the C method. In Sec.~\ref{sec:perturb} the C method is used
to perturbatively calculate the Casimir energy as a power series in
the profile height. Section \ref{sec:algorithm} describes in detail
the numerical algorithm used to calculate Casimir quantities. In
Sec.~\ref{sec:results} the numerical results are explored and compared
to the PFA and its first correction using the DE approach, and the
perturbative approximation. A summary of the work and its main
conclusions are presented in Sec.~\ref{sec:conclusion}. Details of our
calculations are relegated to the appendix.

\section{Scattering Formalism for the Casimir Energy}
\label{sec:scat}
The scattering method has been extensively used for the calculation of
the Casimir forces between the objects with different geometries and
material properties.  In this paper, we use the scattering method to
find the Casimir energy between two planes with a $1-D$ periodic
structure. The method can be easily extended to $2-D$ periodic
structures. In general, the Casimir free energy between two objects
at the temperature T is given by
\begin{equation} \label{eq:energy}
  E = \frac{k_{\mathrm{B}}T A}{2} \sump_{l=0}^\infty
  \int_{\mathbb{B}_1} \!\!\!\! \dif k_\perp
  \ln \det \big( 1 - \mathbb{R}^1\mathbb{U}^{12}
  \mathbb{R}^2\mathbb{U}^{21}\big),
\end{equation}
with $\mathbb{U}$ the translation matrix and $\mathbb{R}^i$ the
scattering matrices of the objects. Both the translation and
scattering matrices depend upon the imaginary Matsubara frequencies
\begin{equation}
  \omega_l = \imath \zeta = \imath \frac{l \pi k_{\mathrm{B}} T}{\hbar}.
\end{equation}
From Eq.~\eqref{eq:energy} for the energy, the Casimir forces or
torques between two objects can be calculated by taking derivatives.
Indeed, the scattering method simplifies the fluctuation-induced
problems by separating the calculation into finding the translation
matrices ($\mathbb{U}$) and scattering matrices ($\mathbb{R}^i$). The
$\mathbb{U}$ matrix corresponds to the way the fluctuations propagate
through the field between the objects and the $\mathbb{R}^i$-matrix
represents the interaction of the object with the fluctuations.  Thus
the information about the distance between the objects is only
contained in the translation matrices.
The elements of the translation and scattering matrices are generally
calculated in a coordinate system appropriate to the geometry of an
object. In the next section we present the vector basis function
suitable for a corrugated system.

\subsection{Vector Basis Functions}
For a periodic system the obvious choice of the vector basis functions
are Block-periodic plane waves
\begin{subequations} \label{eq:basis}
\begin{align}
  \vec{\Psi}^{\text{TE}(\pm)}_{mn} & = 
  \nabla \times \phi^{(\pm)}_{mn}\hat{c}, \\
  \vec{\Psi}^{\text{TM}(\pm)}_{mn} & =
  \frac{1}{\zeta/c} \nabla \times
  \nabla \times \phi^{(\pm)}_{mn}\hat{c},
\end{align}
\end{subequations}
with the $\hat{c}$ vector a constant vector known as a pilot vector
and $\phi^{(\pm)}$ solutions to the scalar Helmholtz equation,
\begin{equation}
\label{eq:helmholtz}
(-\nabla^2 +\zeta^2/c^2)\phi^{(\pm)}=0\,,
\end{equation}
 which are
\begin{equation}\label{eq:scalarbasis}
  \phi^{(\pm)}_{mn}=\exp\big( \imath \vec{K}_{mn} \cdot x_\perp \pm
  \sqrt{ \zeta^2/c^2 +\vec{K}^2_{mn}}z \big),
\end{equation}
These basis functions are recognizable as the simple plane wave vector
functions where the transverse wave-vector has been replaced with the
Block wave-vector $k_\perp \to \vec{K}_{mn}$.  The block wave-vector
can be written as
\begin{equation}\label{eq:KGvecs}
  \vec{K}_{mn}=k_\perp + \vec{G}_{mn},
\end{equation}
with $k_\perp$ a continuous wave-vector that only takes on values in
the first Brillouin zone and $\vec{G}_{mn}$ a discrete lattice vector given by
\begin{equation}\label{eq:Kvecs}
  \vec{G}_{mn}=\vec{b_1}m+\vec{b_2}n
\end{equation}
where $\vec{b}_1$ and $\vec{b}_2$ are inverse lattice vectors of
the periodic system, and $m$ and $n$ are integers.  Using the
aforementioned basis, we can now define the translation and scattering
matrices.
\subsection{Translation Matrix}
Because the vector basis functions are essentially plane waves, the
translation matrix is simply
\begin{multline}\label{eq:U}
  \big(\mathbb{U}^{12})^{p,p'}_{mn,m'n'}=
  \delta_{p,p'}\delta_{mn,m'n'}\\
  \times \exp\big(\imath{K_{mn}}\cdot \vec{b}_\perp
    -\sqrt{\zeta^2/c^2+\vec{K}^2_{mn}}d\big) ,
\end{multline}
with $\vec{b}_\perp$ a in plane displacement, and $d$ a perpendicular
separation. It should be noted that for a fixed separation $d$, the
$\mathbb{U}$ matrix is exponentially suppressed for large imaginary
frequency $\zeta$. In addition for fixed separation $d$ and fixed
imaginary frequency $\zeta$, the elements of the translation matrix
are exponentially suppressed in $m$ and $n$. Both of these features
are needed for the Casimir quantities to converge (both in the
frequency integral, and as a function of matrix size) and for making
the evaluation of relevant matrices numerically efficient.

\subsection{Scattering Matrix}
\begin{figure}
  \includegraphics[width=80mm]{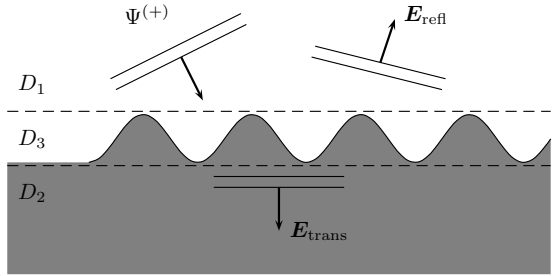}
  \caption{\label{fig:scat} Scattering problem for surface relief
    grating.}
\end{figure}
To obtain the scattering matrix, we divide the space into three
regions as illustrated in Fig.~\ref{fig:scat}: region $D_1$ completely
above the periodic surface, region $D_2$ completely below the periodic
surface, and region $D_3$ including the periodic surface. We now
consider an incident wave ($\vec{\Psi}^{(+)}_{mn}$) is scattered by
the surface in region $D_3$, and is either reflected back into region
$D_1$ or transmitted into region $D_2$
\begin{equation}
  \vec{E}^p_{mn}=\begin{cases}
  \vec{\Psi}^{p(+)}_{mn}+
  \vec{E}^p_{mn,\text{refl}}& 
  \text{in $D_1$},\\
  \vec{E}^p_{mn,\text{trans}}& 
  \text{in $D_2$}.
  \end{cases}
\end{equation}
The reflected or transmitted field can be written as a sum over the
complete set of vector basis functions. Furthermore because of the
boundary conditions at infinity the reflected wave in region $D_1$
only contains exponentially dying wave and the transmitted wave in
region $D_2$ only contains exponentially growing waves (dying in the
negative $z$ direction). Because the system is periodic, the
Floquet-Bloch theorem states that the solution must be pseudo-periodic
(periodic with a phase factor). The reflected and transmitted fields
can then be completely written as
\begin{subequations} \label{eq:reflect}
\begin{align}
  \vec{E}^p_{mn,\text{refl}}& = \sum_{p'}
  \sum_{n'm'}\mathbb{R}_{mn,m'n'}^{p,p'} \vec{\Psi}^{p'(-)}_{m'n'},
  \\ \vec{E}^p_{mn,\text{trans}}& = \sum_{p'}
  \sum_{n'm'}\mathbb{T}_{mn,m'n'}^{p,p'} \vec{\Psi}^{p'(+)}_{m'n'}.
\end{align}
\end{subequations}
This is known as the Rayleigh expansion with the matrix elements of
the $\mathbb{R}$ and $\mathbb{T}$ matrices called the
Rayleigh coefficients.

For the remainder of this work many simplifications will be performed to
make the derivations more tractable, and the results easier to
analyze. We will consider a two parallel 1-D periodic systems
alligned along the axis of corrugations made of perfectly conducting
materials at zero temperature.

\subsection{1-D corrugation perfect metal at zero temperature}

We consider 1-D corrugations that are translationally invariant in the
$y$ direction. A natural direction for the pilot vector in
Eqs.~\eqref{eq:basis} is then $\hat{c}=\hat{y}$.  The full electric
field can then be rewritten in terms of two scalar fields, the TM and
TE modes defined as
\begin{subequations}
\begin{align}
  E_y &= \int \!\! \dif k_y \; e^{\imath k_y y} f^{\text{TM}}, \\
  H_y &= \int \!\! \dif k_y \; e^{\imath k_y y} f^{\text{TE}},
\end{align}
\end{subequations}
where both $f$ fields satisfy the Helmholtz equation
\begin{equation}\label{eq:fdifeq}
  \big[-\partial_x^2-\partial_z^2+\kappa^2\big]f=0,
\end{equation}
with $\kappa^2=\zeta^2/c^2+k_y^2$. The scattering problem can be written
\begin{equation}\label{eq:ftot}
  f^p_{m,\text{tot}} = \phi^{(+)}_{m}+ f^{p}_{m,\text{refl}} 
  \qquad\text{in $D_1$},
\end{equation}
where the incident wave $\phi^{(+)}$ is the Fourier transform of
scalar basis function in Eq.~\eqref{eq:scalarbasis}
\begin{equation}\label{eq:fscatprob}
  \phi^{(\pm)}_m(x,z) =
  \exp\big(\imath \vec{K}_m x \pm 
  \sqrt{\kappa^2 + \vec{K}^2_m} z\big).
\end{equation}
For perfect electrical conductors the boundary conditions reduce to
Dirichlet and Neumann boundary conditions,
\begin{subequations}\label{eq:fbc}
\begin{align} \label{eq:ftmbc}
  f^{\text{TM}}(x,z) \hspace{-5.5ex} 
  \operatorname*{\big|}_{\hspace{5ex} z=h(x)} \hspace{-5.5ex}
  &=0,\\ \label{eq:ftebc}
  \hat{n} \cdot \nabla f^{\text{TE}}(x,z) \hspace{-5.5ex}
  \operatorname*{\big|}_{\hspace{5ex} z=h(x)} \hspace{-5.5ex}
  &=0.
\end{align}
\end{subequations}
The zero temperature condition will change the sum over Matsubara
frequencies given in Eq.~\eqref{eq:energy} to an integral over imaginary
frequency.  The Casimir energy per unit length between two 1-D perfect
metal corrugations can then be written as
\begin{multline}\label{eq:en1}
  \frac{E}{L_y} = \frac{\hbar c L_x}{8 \pi^2}
  \int_0^\infty \!\!\!\!\! \kappa \dif \kappa
  \int_{-\pi/L_x}^{\pi/L_x} \!\!\!\!\!\!\!\!\! \dif k_x \\
  \sum_p \ln \det\big( 1-
  \mathbb{R}^{1p}\mathbb{U}^{12}\mathbb{R}^{2p}\mathbb{U}^{21}
  \big),
\end{multline}
where the sum over the polarization $p$ contains the TE and TM
modes.  To calculate the Casimir energy using Eq.~\eqref{eq:en1}, we
need to find the scattering matrix $\mathbb{R}$ that we obtain in the
next section using the C-method.

\section{C-Method}
\label{sec:Cmeth}
The C method was developed as an efficient numerical method for
calculating the Rayleigh coefficients for surface relief
gratings \cite{Chandezon:1980a}. In this section, we describe the C
method for simple 1-D perfectly conducting boundary conditions. For
other boundary conditions see Ref.~\cite{Chandezon:2002a} and
references therein.  In what follows we present the C method for the
corrugated system, which involves an explicit change of variable to
remove the $z$ dependence, the direction perpendicular to the mean
surface of grating, followed by a Fourier transform in the $x$ and $y$
coordinates. This procedure expresses the Helmholtz equation,
Eq.~\ref{eq:helmholtz}, as a quadratic eigenvalue problem more
amenable to numerical solutions.

We start with the following change of variable 
\begin{equation}
  \{u,v,w\} = \{x,y,z-h(x)\}.
\end{equation}
While this change of variable will have the effect of explicitly
removing the $z$ dependence from the boundary condition, it will
introduce the gradient of the profile function into the Helmholtz
equation through the partial derivatives
\begin{equation}\label{eq:partial_expansion}
  \partial_x f(x,z) \to
  \big(\partial_u - (\partial_u h) \partial_w\big)f(u,w),
\end{equation}
In the next step, we write the profile function $h$ and the $f$ fields
in a Fourier series
\begin{subequations}\label{eq:fourier}
\begin{align} \label{eq:hudef}
  h(u) &= 
  \sum_{m}e^{\imath \vec{G}_{m} u}h_{m}, \\ \label{eq:fuwdef}
  f(u,w) &= 
  \sum_{m}e^{\imath \vec{K}_{m} u}f_{m}(w),
\end{align}
\end{subequations}
where $\vec{G}_m$ and $\vec{K}_m$ are the inverse lattice and Block
vectors defined in Eq.~\eqref{eq:KGvecs}. For the 1-Dimensional
periodic profiles considered the vectors can be explicitly written
\begin{equation}\label{eq:G}
  \vec{K}_m = k_x + \vec{G}_m \quad\text{and}\quad
  \vec{G}_m = \frac{2\pi}{L_x}m,
\end{equation}
where $L_x$ is the period of the profile. The partial derivatives of
$f$ based on Eqs.~\eqref{eq:partial_expansion} and \eqref{eq:fourier}
then yield
\begin{equation}\label{eq:partial1}
  \partial_u f(u,w) = \sum_m e^{\imath \vec{K}_m u}
  \big(\imath \vec{K}_m f_m(w)\big),
\end{equation}
and
\begin{equation}\label{eq:partial2}
  (\partial_u h) \partial_w f(u, w) =
  \sum_{m,m'} e^{\imath (\vec{K}_m + \vec{G}_{m'}) u}
  \big(\imath \vec{G}_{m'} h_{m'} \partial_w f_m(w)\big),
\end{equation}
By combining
$\vec{K}_m+\vec{G}_{m'} = \vec{K}_{m+m'}$, and changing the
variable $m' \to n-m$, Eq.~\eqref{eq:partial2} becomes
\begin{equation} \label{eq:partial3}
  (\partial_u h) \partial_w f(u, w) =
  \sum_{n} e^{\imath \vec{K}_n u} 
  \sum_m \imath \vec{G}_{n-m} h_{n-m} \partial_w f_m(w).
\end{equation}
Using Eq.~\eqref{eq:partial1} and \eqref{eq:partial3},
Eq.~\eqref{eq:partial_expansion} can be written in the following
compact form
\begin{equation}
  \partial_x f(x,z) \to 
  \sum_m e^{\imath \vec{K}_m u}
  \Big( \imath \big(\mat{K} - \mat{Gh} \partial w \big) \cdot \vec{f}(w)
    \Big)_m ,
\end{equation}
such that
\begin{subequations} \label{eq:gh}
\begin{equation}\label{eq:fvec}
  \big(\vec{f}(w)\big)_m \equiv f_m(w),
\end{equation}
and $\mat{K}$ and $\mat{Gh}$ are matrices with elements defined by
\begin{align} \label{eq:Kmat}
  (\mat{K})_{m,m'} &\equiv 
  \delta_{m,m'}\vec{K}_{m}, \\ \label{eq:Ghmat}
  (\mat{Gh})_{m,m'} &\equiv
  \vec{G}_{(m-m')}h_{(m-m')}.
\end{align}
\end{subequations}

Separating out the Fourier modes, the Helmholtz equation can now be
written as an infinite system of ordinary differential equations
\begin{equation} \label{eq:diff}
  \big( (\mat{K} -\mat{Gh}\partial_w)^2 - \mat{I}\partial_w^2 +
  \mat{I}\kappa^2 \big)\cdot\vec{f}(w)=0,
\end{equation}
whose solution is assumed to have an exponential form
\begin{equation} \label{eq:sol1}
  \vec{f}(w) = \vec{V} e^{\lambda w},
\end{equation}
with eigenvalue $\lambda$ and eigenvector $\vec{V}$.
Upon substitution of Eq.~\eqref{eq:sol1} into
Eq.~\eqref{eq:diff}, we obtain a quadratic eigenvalue problem for the
eigenvalues and eigenvectors,
\begin{equation}\label{eq:qep}
  \lambda^2_q 
  \big(\mat{A}_2 - \mat{I}\big)
  \cdot \vec{V}_q  - 
  \lambda_q 
  \mat{A}_1 \cdot \vec{V}_q  + 
  \mat{A}_0 \cdot
  \vec{V}_q= 0,
\end{equation}
with
\begin{subequations}\label{eq:Amats}
\begin{align} \label{eq:A2mat}
  \mat{A}_2 = &
  \mat{Gh}\cdot\mat{Gh},\\ \label{eq:A1mat}
  \mat{A}_1 = &
  \mat{K}\cdot\mat{Gh}+\mat{Gh}\cdot\mat{K},\\ \label{eq:A0mat}
  \mat{A}_0 = &
  \mat{I} \kappa^2 + \mat{K}\cdot\mat{K}.
\end{align}
\end{subequations}
The general solution to the full wave equation can now be written by
combining Eqs.~\eqref{eq:fuwdef}, \eqref{eq:fvec}, and \eqref{eq:sol1}
\begin{equation} \label{eq:sol}
  f(u,w) = \sum_m e^{\imath K_m u} \sum_q c_q \big(\vec{V}_q\big)_m
  e^{\lambda_q w},
\end{equation}
with $q$ indexing the solutions to the quadratic eigenvalue problem,
and $c_q$ undetermined coefficients.
The undetermined coefficients $c_q$ can be
found by applying the boundary conditions given in \eqref{eq:fbc}.

\subsection{Boundary Conditions}
\label{sec:bc} 
To apply the boundary conditions in Eqs.~\eqref{eq:fbc} to the total
field $f_{\text{tot}}$
\begin{equation}
  f^p_{m, \text{tot}}(u,w) = \phi^{(+)}_m(u,w) 
  + f^p_{m, \text{refl}}(u,w),
\end{equation}
we need to write the incident and reflected waves in a Bloch series.

The incident wave is simply an exponentially growing (dying in the
negative $z$ direction) plane wave basis function indexed by
$m$. After a change in variables to the $\{u,v,w\}$ coordinates, the
plane wave can be written
\begin{equation} \label{eq:phi+}
  \phi^{(\pm)}_m(u,w) = 
  e^{\imath \vec{K}_m u \pm 
  \widetilde{\lambda}_m (w + h(u))},
\end{equation}
with $\widetilde{\lambda}_m$ the Rayleigh wavenumber
\begin{equation}\label{eq:ltdef}
  \widetilde{\lambda}_m = \sqrt{\kappa^2 + \vec{K}_m^2}.
\end{equation}
Note the change in variable $z \to w + h(u)$ introduces an
additional $u$ dependence such that Eq.~\eqref{eq:phi+} is not
strictly a Fourier series in $u$. However, we can still expand the
incident wave in a Fourier series 
\begin{equation} \label{eq:fincident}
  \phi^{(\pm)}_m(u,w) =
  \sum_{m'} e^{\imath \vec{K}_{m'} u}
  \mathcal{L}^{(\pm)}_{m m'} e^{\pm\widetilde{\lambda}_m w},
\end{equation}
where the $\mathcal{L}$ term are the Fourier coefficients,
\begin{equation}\label{eq:Ldef}
  \mathcal{L}^{(\pm)}_{m m'} = 
  \int \!\! \dif u e^{-\imath G_{m'-m} u \pm \widetilde{\lambda}_m h(u)}.
\end{equation}

Further, the reflected wave can be written as in Eq.~\eqref{eq:sol} in
terms of the eigenvalues and eigenvectors of the quadratic eigenvalue
problem
\begin{equation}\label{eq:c_uw}
  f^p_{m,\text{refl}}(u,w)=\sum_{m'}  e^{\imath \vec{K}_{m'} u}
  \sum_{q\in\{\lambda_-\}}
  c^p_{mq}\big(\vec{V}_q\big)_{m'}e^{\lambda_q w}.
\end{equation}
It should be noted that because the reflected wave must go to zero in
the limit $w\to +\infty$ we need only to use the set of eigenvalues
$\{\lambda_-\} = \{\lambda_q | Re(\lambda_q)<0\}$ and their associated
eigenvectors. The quantity $m$ in Eq.~\eqref{eq:c_uw} corresponds to the
Fourier index of the incident wave, and the $p$ index labels the mode
as either TM or TE.

Now, inserting Eqs.~\eqref{eq:fincident} and \eqref{eq:c_uw} in the
boundary condition Eqs.~\eqref{eq:ftmbc} and \eqref{eq:ftebc}, and
separating out the modes, we find a system of equations for the
unknown coefficients $c^p_{mq}$ written in matrix form as
\begin{equation}\label{eq:Fmateq}
  \sum_{q \in \{\lambda_-\}} \mat{F}^p_{m'q} c^{p}_{mq} = \vec{b}^p_{m m'},
\end{equation}
where $\mat{F}^p$ and $\vec{b}^p$ are given explicitly for the TM and
TE modes below.

For the TM mode, the total field $f_{\text{tot}}$ at $w=0$ obeys
Dirichlet boundary conditions
\begin{equation}
  f^{\text{TM}}_{\text{tot}}(u,0)=0.
\end{equation} 
In this case the $\mat{F}$ matrix is simply the matrix of eigenvectors
\begin{equation}\label{eq:FTM}
  \mat{F}^{\text{TM}}_{m' q} =
  \big(\vec{V}_q\big)_{m'},
\end{equation}
and the $\vec{b}$ vectors are simply the Fourier coefficients of
the incident wave
\begin{equation}\label{eq:bTM}
  \vec{b}^{\text{TM}}_{m m'}
  = -\mathcal{L}^{(+)}_{m m'}.
\end{equation}
The TE mode obeys Neumann boundary conditions given in Eq.~\eqref{eq:ftebc}. In the $\{u,v,w\}$ coordinates
Eq.~\eqref{eq:ftebc} becomes
\begin{equation}
  \big(-h'\partial_u +
  (1+h^{\prime 2})\partial_w\big) 
  f^{\text{TE}}_{\text{tot}}(u,w)
  \hspace{-3ex}
  \operatorname*{\big|}_{\hspace{3ex} w=0}
  \hspace{-3ex} = 0,
\end{equation}
with $h'$ the first derivative of the profile function $h(u)$ with
respect to $u$. After separating out the Fourier modes, the $\mat{F}$
matrix is
\begin{equation}\label{eq:FTE}
  \mat{F}^{\text{TE}}_{m' q} =
  \Big(\big( \mat{Gh}\cdot\mat{K}
  + \lambda_q (\mat{I}-\mat{Gh}\cdot\mat{Gh})
  \big)\cdot \vec{V}_q \Big)_m ,
\end{equation}
and the $\vec{b}$ vector is
\begin{multline}\label{eq:bTE}
  \vec{b}^{\text{TE}}_{m m'} = -\sum_{m''}
  \big( \mat{Gh}\cdot\mat{K}+ \\ \widetilde{\lambda}_m 
  (\mat{I}-\mat{Gh}\cdot\mat{Gh})
  \big)_{m'm''} \mathcal{L}^{(+)}_{mm''}.
\end{multline}
with matrices $\mat{Gh}$, $\mat{K}$, and the vector
$\mathcal{L}^{(+)}$ defined in Eqs~\eqref{eq:Kmat},\eqref{eq:Ghmat},
and \eqref{eq:Ldef}, respectively. Utilizing these
expressions it is possible to solve for the unknown coefficients
$c^p_{mq}$, which in turn gives the exact form of the field $f$ for
the scattering problem.

\subsection{Identifying Rayleigh Coefficients}
\label{sec:rayleigh}
In order to find the Rayleigh coefficients we must compare the
expression for the reflected wave in Eq.~\eqref{eq:c_uw} to that in
region $D_1$ using the Rayleigh expansion as given in
Eq.~\eqref{eq:reflect}. An expansion analogous to
Eq.~\eqref{eq:reflect} for a 1-D corrugations in the $\{u,v,w\}$
coordinate system yields
\begin{equation}\label{eq:fray}
  f^{p}_{m,\text{refl}}(u,w) = 
  \sum_{m'}\mathbb{R}^{p}_{mm'}
  \phi^{(-)}_{m'}(u,w).
\end{equation}
Using the Fourier expansion of the basis functions as given in
Eq.~\eqref{eq:fincident}, we write the full reflected wave in Eq.~\eqref{eq:fray} as
\begin{equation}\label{eq:R_uw}
  f^{p}_{m,\text{refl}}(u,w) = \sum_{m''}
   e^{\imath \vec{K}_{m''} u} \sum_{m'}\mathbb{R}^{p}_{mm'}
  \mathcal{L}^{(-)}_{m'm''}e^{-\widetilde{\lambda}_{m'} w},
\end{equation}
with $\widetilde{\lambda}$ the Rayleigh wavenumber defined in
Eq.~\eqref{eq:ltdef} and $\mathcal{L}$ the Fourier coefficients of the
incident wave given in Eq.~\eqref{eq:Ldef}.

Equation \eqref{eq:R_uw} corresponds to the Rayleigh expansion in the
$\{u,v,w\}$ coordinates. The Rayleigh coefficients can be obtained by
matching Eq.~\eqref{eq:sol} with Eq.~\eqref{eq:R_uw} term by term,
\begin{equation} \label{eq:Ray}
  \mathbb{R}^p_{mm'} = c^p_{mq(m')}\frac{\big(\vec{V}_{q(m')}\big)_{m'}}
         {\mathcal{L}^{(-)}_{mm'}},
\end{equation}
where $q(m')$ is the index of the eigenvalue that matches with the
$m^{\prime\text{th}}$ Fourier index. Note that the eigenvectors
$\vec{V}_q$ and eigenvalues $\lambda_q$ are determined by the
quadratic eigenvalue problem in Eq.~\eqref{eq:qep}. The $c^p_{mq}$
coefficients are determined by solving the linear system in
Eq.~\eqref{eq:Fmateq}.  Equations \eqref{eq:qep},
\eqref{eq:Fmateq}, and \eqref{eq:Ray} can be used to numerically
calculate the Rayleigh coefficients, and through them the Casimir
energy. However, it is possible to obtain analytical results in
certain limits, which we present in the next section.

\section{Small Amplitude Perturbation}
\label{sec:perturb}
In the limit of small amplitude surface relief gratings, perturbation
theory can be used to analytically obtain the Casimir energy as a
series in the height profile $h(x)$. Using the C method we can obtain
a perturbative expression for the Rayleigh coefficients $\mathbb{R}$,
which requires perturbative expressions for the eigenvectors
$\vec{V}_q$ and eigenvalues $\lambda_q$.

Solving the quadratic eigenvalue problem given in Eq.~\eqref{eq:qep}
perturbatively, we expand the matrices $\mat{A}_1$ and $\mat{A}_2$ up
to the order $\mathcal{O}(h^2)$ and $\mathcal{O}(h)$ respectively. In
addition, we set the eigenvalues $\lambda = \sum_i \lambda^{(i)}$ and
eigenvectors $\vec{V} = \sum_i \vec{V}^{(i)}$, where the
$(i)^{\text{th}}$ term is of order $\mathcal{O}(h^i)$. Grouping the
terms together by powers of $h$, we find the zeroth order equation
\begin{equation} \label{eq:zero}
  -\big(\lambda^{(0)}_m\big)^2 \vec{V}^{(0)}_m 
  + \mat{A}_0 \cdot \vec{V}^{(0)}_m =0.
\end{equation}
The $\mat{A}_0$ matrix is diagonal and thus the
eigenvectors are given by Kronecker delta functions,
  $\big( \vec{V}^{(0)}_m \big)_{m'} = \delta_{m m'}$.
The eigenvalues are the square root of the diagonal elements of
the $\mat{A}_0$ matrix. In order to get the exponentially dying
components, we consider only the negative eigenvalues
\begin{equation} \label{eq:lam0}
  \lambda^{(0)}_m = - \sqrt{\kappa^2 + \vec{K}_m^2} = 
  - \widetilde{\lambda}_m.
\end{equation}
This is the same as the negative Rayleigh wavenumber given by
Eq.~\eqref{eq:ltdef}. To the zeroth order, the reflected waves are
just plane waves.  

The first two corrections to the eigenvalues are
\begin{align}\label{eq:lam1}
  \lambda^{(1)}_m & = 0, \\ \label{eq:lam21}
  \lambda^{(2)}_m & = -\widetilde{\lambda}_m \vec{K}_m 
  \sum_{m'} |h_{m-m'}|^2 \vec{G}_{m'-m}.
\end{align}
The derivations of Eqs.~\eqref{eq:lam1} and \eqref{eq:lam21} are given
in Appendix \eqref{app:ev}. The careful examination of
Eq.~\eqref{eq:lam21} shows that the $m'$ sum is exactly zero for $m=0$
and very small for $m$ near zero. Note that the $\vec{G}_{m-m'}$ term
is exactly zero for $m=m'$ and for $m\ne m'$, $|h_{m-m'}|^2$ is even
in $m'$ around $m$ while $\vec{G}_{m'-m}$ odd. Thus, the sum of the
$m+m'$ and $m-m'$ terms is exactly zero. For finite matrices where $m$
ranges from $-M$ to $M$ the cancellation will only occur exactly for
$m=0$. For $m$ near zero, if $M \gg m$ then most of the $m'$ terms
will cancel, leaving terms where $|m'| \sim M$. If the Fourier
coefficients $h_m$ decay fast enough then the second order correction
is negligible for $m$ near 0.

The first and second order corrections to the zeroth order
eigenvectors are 
\begin{align} \label{eq:V1}
  \big(\vec{V}^{(1)}_m\big)_{m'} &=
  - \widetilde{\lambda}_m h_{m'-m}, \\ \nonumber
  \big(\vec{V}^{(2)}_m\big)_{m'} &=
  \frac{\widetilde{\lambda}_m^2}{2}
  \sum_{m''} h_{m'-m''}h_{m''-m} -  \\ \label{eq:V2}
  & \quad \frac{\widetilde{\lambda}_m^2}{2}
  \sum_{m''} h_{m'-m''}h_{m''-m} 
  \frac{\vec{G}_{m+m'-2m''}}{\vec{G}_{m'-m}}.
\end{align}
Similar to the situation with the second order eigenvalues, the second
term in Eq.~\eqref{eq:V2} can be shown to be negligible for $m$ near
zero.

We now use Eqs.~\eqref{eq:Fmateq} and \eqref{eq:Ray} to find a perturbative expansion for the Rayleigh
coefficients. The expressions for the $F$
matrices in Eq.~\eqref{eq:Fmateq} can be found using the perturbative expansions for the
eigenvalues and eigenvectors in Eqs.~\eqref{eq:FTM} and
\eqref{eq:FTE}. However we still need a pertubative expansion for the
$\mathcal{L}$ term used in Eqs.~\eqref{eq:bTM}, \eqref{eq:bTE}, and
\eqref{eq:Ray}. The $\mathcal{L}^{(\pm)}$ terms are the Fourier
coefficients of an incident (+) or a scattered (-) plane wave in the
$\{u,v,w\}$ coordinate system.  Using a series expansion in powers of
the height profile $h$, $\mathcal{L}^{(\pm)}_{mm'} = \sum_i
\mathcal{L}^{(\pm)(i)}_{mm'}$, each term can be identified as the
Fourier coefficients of powers of the profile function given in
Eq.~\eqref{eq:Ldef}. The first three terms are
\begin{subequations} \label{eq:Lexp}
\begin{align}
  \mathcal{L}^{(\pm)(0)}_{mm'}&=\delta_{mm'},\\ 
  \mathcal{L}^{(\pm)(1)}_{mm'}&= 
  \pm \widetilde{\lambda}_m h_{m'-m}, \\ 
  \mathcal{L}^{(\pm)(2)}_{mm'} & =
  \frac{\widetilde{\lambda}_m^2}{2} \sum_{m''}h_{m'-m''}h_{m''-m}.
\end{align}
\end{subequations}
It should be noted that the zeroth, first, and second order expansions
of the eigenvectors in the perturbative expansion exactly match with
the first three terms of a perturbative expansion of the plane wave in
the $\{u,v,w\}$ coordinate system given by
Eq.~\eqref{eq:fincident}. From this expansion it is possible to make
the identification
\begin{equation}\label{eq:VeqL}
  \big(\vec{V}_m^{(i)}\big)_{m'}=
  \mathcal{L}^{(-)(i)}_{mm'},
\end{equation}
through second order. {Note that the zeroth order eigenvalue is
exactly equal to the Rayleigh wavenumber for the plane wave as given in Eq.~\eqref{eq:lam0}, and the first
two corrections are zero, see Eqs.~\eqref{eq:lam1} and \eqref{eq:lam21}.

The equality in equation \eqref{eq:VeqL} between the basis functions
identified through the C-method and simple plane waves seems to imply
the Rayleigh hypothesis, which consider the solution to the scattering
problem can be written in terms of only the exponentially dying waves
even inside the grooves in region $D_3$. Note that in
Refs.~\cite{Petit:1966a,Apeltsin:1985a} it is shown that the Rayleigh
hypothesis is valid for certain profiles with small enough height
amplitudes. We emphasize that the perturbative expansion presented in
this section is performed under the assumption that the maximum
profile height is smaller than all other length scales in the
system. Thus we expect that the equality presented in
Eq.~\eqref{eq:VeqL} would be true to all orders.

Inserting Eq.~\eqref{eq:VeqL} into Eq.~\eqref{eq:Ray}, we find the
Rayleigh coefficients are exactly given by the undetermined
coefficients $\mathbb{R}=c$. Using the perturbative expressions for
the eigenvalues and eigenvectors, Eqs.\eqref{eq:lam0}-\eqref{eq:V2}},
it is possible to solve Eq.~\eqref{eq:Fmateq} perturbatively for the
Rayleigh coefficients. The expressions for the $TM$ modes are
\begin{subequations}\label{eq:Rtmexp}
\begin{align}\label{eq:Rtm0}
  \mathbb{R}^{\text{TM}(0)}_{mm'} & = - \delta_{mm'}, \\ \label{eq:Rtm1}
  \mathbb{R}^{\text{TM}(1)}_{mm'} &= 
  - 2 \widetilde{\lambda}_m h_{m'-m}, \\ \label{eq:Rtm2}
  \mathbb{R}^{\text{TM}(2)}_{mm'} &= 2 
  \widetilde{\lambda}_m \sum_{m''} 
  \widetilde{\lambda}_{m''}h_{m'-m''}h_{m''-m},
\end{align}
\end{subequations}
and for $TE$ modes are
\begin{subequations}\label{eq:Rteexp}
\begin{align}\label{eq:Rte0}
  \mathbb{R}^{\text{TE}(0)}_{mm'} &= \delta_{mm'},\\ \label{eq:Rte1}
  \mathbb{R}^{\text{TE}(1)}_{mm'} &=
  2 \frac{\widetilde{\lambda}_{mm'}^2}
  {\widetilde{\lambda}_{m'}}h_{m'-m},\\ \label{eq:Rte2}
  \mathbb{R}^{\text{TE}(2)}_{mm'} &=
  2 \sum_{m''}
  \frac{\widetilde{\lambda}_{mm''}^2\widetilde{\lambda}_{m'm''}^2}
       {\widetilde{\lambda}_{m''}\widetilde{\lambda}_{m'}}
       h_{m'-m''}h_{m''-m},
\end{align}
\end{subequations}
where the $\widetilde{\lambda}_{mm'}$ is a modified Rayleigh
wave-vector given explicitly as
\begin{equation}
  \widetilde{\lambda}_{mm'}=\sqrt{\kappa^2 + \vec{K}_m \vec{K}_m'}.
\end{equation}
A more detailed derivation of Rayleigh coefficients is presented in
Appendix \ref{app:refl}.

Inserting Eqs.~\eqref{eq:Rtmexp} and \eqref{eq:Rteexp} into
Eq.~\eqref{eq:energy}, we can obtain the perturbative expansion for
the Casimir energy in powers of the grating profile $h$. For a single
grating above a flat sheet, the zeroth order term gives the expression
for the Casimir energy between two parallel plates,
which is expected as the zeroth order reflection coefficients
correspond to those for flat plates. The first order correction is
\begin{equation}\label{eq:enp1}
  \frac{E^{(1)}}{L_y L_x} = -\frac{\pi^2 \hbar c}{240}
  \frac{h_0}{d^4},
\end{equation}
with $h_0$ the zeroth Fourier mode of the height profile, also the
average height of the grating. Equation \refeq{eq:enp1} is equal to
zero if we define the profile to have zero average height (such as a
sinusoidal grating). The second order correction to the energy is
\begin{multline}\label{eq:enp2}
  \frac{E^{(2)}}{L_y L_x} = -\frac{\pi^2 \hbar c}{240}
  \sum_m \frac{|h_m|^2}{d^5} \\ \Big(
  g_{TM}\big(\tfrac{4 \pi m d}{L_x}\big)+
  g_{TE}\big(\tfrac{4 \pi m d}{L_x}\big)
  \Big),
\end{multline}
where the $g_{TM}$ and $g_{TE}$ are integral expressions given
explicitly in Eqs.~\ref{eq:gterms} in App.~\ref{app:en}. The complete
details of the derivation of perturbative energies and the lateral
Casimir forces are given in App.~\ref{app:en}. It is important to note
that the expressions for the energy and lateral force given in
Eqs.~\eqref{eq:enp2} and \eqref{eq:Flatteral} exactly match the
previous results obtained in Refs.~\cite{Emig:2001dx, Emig:2002xz,
  CaveroPelaez:2008tj, CaveroPelaez:2008rt}.


\section{Numerical Method}
\label{sec:algorithm}
Here we employ the C method described in previous sections to calculate
the Casimir energy through Eq.~\eqref{eq:en1}.
The integrand in Eq.~\eqref{eq:en1} depends on the height profile
$h(x)$ or its Fourier components $h_m$, the combined wave-vector
$\kappa$, the wave-vector in the $x$ direction $k_x$, and the maximum
Forier mode $M$. All the relevant matrices will then be of size
$N \times N$, with $N=2M+1$. As an example we assume a sinusoidal
profile with $h(x)=a\sin(2\pi x)$.  To obtain eigenvalue eigenvector
pairs $\{\lambda,\vec{V}\}$ presented in Eq.~\eqref{eq:qep}, we need
to generate matrices $\mat{A}_0$, $\mat{A}_1$, and $\mat{A}_2$.  For
the sinusoidal profile function the Fourier components are trivially
found, $h_m = -\imath a/2$.  The matrix elements of the $\mat{Gh}$
matrix (see Eqs.~\eqref{eq:G} and \eqref{eq:Ghmat}) are
\begin{equation}
  \mat{Gh}_{mm'} = -\imath \pi a \delta_{m,m'\pm1},
\end{equation}
from which we can find 
\begin{align}
  (\mat{A}_2)_{mm'} &= \pi^2 a^2\big( -2 \delta_{m,m'} +
  \delta_{m,\pm N}\delta_{m',\pm N} -
  \delta_{m,m'\pm2} \big),\\
  (\mat{A}_1)_{mm'} &= -2\imath \pi a \big(k_x + \pi(m+m')\big) 
  \delta_{m,m'\pm1},\\
  (\mat{A}_0)_{mm'} &= \kappa^2+(k_x-2\pi m)^2 \delta_{m,m'}.
\end{align}

\begin{figure}[htb]
\includegraphics[width=85mm]{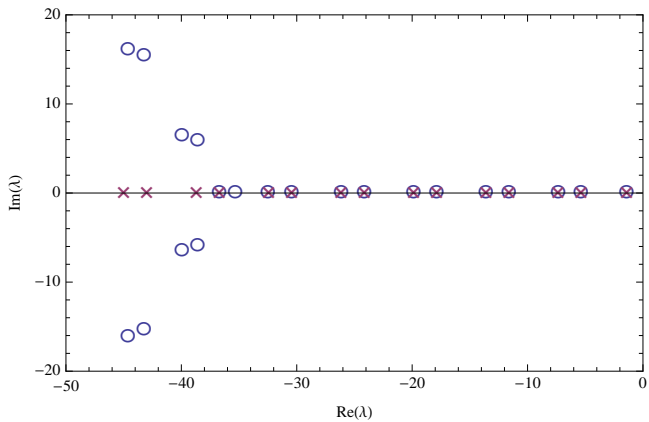}
\caption{\label{fig:eigenvalues} A scatter plot of the eigenvalues
  $\{\lambda_-\}$ (circles) in the complex plane. The problem
  corresponds to the sinusoidal grating $h(x)= 0.1 \sin(2\pi x)$,
  $\kappa = 1$, and $k_x=1$. The matrix size is $21\times21$ with the
  Fourier index $m$ ranging from $-10, 10$. The first 15 negative
  Rayleigh wavenumbers (crosses) defined in Eq.~\eqref{eq:ltdef} are
  also shown on the plot. The figure shows that 11 of the 21
  eigenvalues match well with the Rayleigh wavenumbers.}
\end{figure}

For a given $k_x$, $\kappa$ and a finite $N\times N$ matrix, the
quadratic eigenvalue problem in Eq.~\eqref{eq:qep} will yield $2N$
eigenvalues and eigenvectors that can be obtained numerically. The
standard method is to recast the quadratic eigenvalue problem into a
larger (generalized) eigenvalue problem with a $2N\times2N$
matrix given in block form as
\begin{equation}\label{eq:geneigen}
  \begin{pmatrix} 0 & \mat{I} \\ -\mat{A}_0 & \mat{A}_1 \end{pmatrix}
  \vec{\mathcal{V}} =
  \lambda 
  \begin{pmatrix} \mat{I} & 0 \\ 0 & \mat{A}_2-\mat{I} \end{pmatrix}
  \vec{\mathcal{V}},
\end{equation}
where we can identify the $2N\times 1$ vector $\vec{\mathcal{V}}$ as
constructed from block components of $\vec{V}$ and $\lambda
\vec{V}$. The eigenvalues and eigenvectors for the generalized
eigenvalue problem in Eq.~\eqref{eq:geneigen} are found numerically,
and the first $N$ components of the $\vec{\mathcal{V}}$ eigenvector
are the same as the components of the $\vec{V}$ eigenvector of the
corresponding quadratic eigenvalue problem. We sort the eigenvalue,
eigenvector pairs into those with eigenvalues with positive real part
$\{\lambda_+,\vec{V}_+\}$, and with negative real parts
$\{\lambda_-,\vec{V}_-\}$. We label the negative pairs with an index
$q$ that will run from 1 to $N$. Figure \ref{fig:eigenvalues} shows a scatter plot of the numerically obtained eigenvalues (circles) on
the complex plane. Using the $\{\lambda_-,\vec{V}_-\}$ eigenvalue and
eigenvector pairs and Eqs.~\eqref{eq:FTM} and \eqref{eq:FTE}, we are
now able to generate the $\mat{F}^{\text{TM}}$ and
$\mat{F}^{\text{TE}}$. Our goal is solve to the matrix equation
$\mat{F} c = \vec{b}$ given in Eq.~\eqref{eq:Fmateq} for both $TM$-
and $TE$-polarizations to obtain the $c$ coefficients which later will
be used to obtain the Rayleigh coefficients.

The target vectors $\vec{b}^{\text{TM}}_q$ and $\vec{b}^{\text{TE}}_q$
given in Eqs.~\eqref{eq:bTM} and \eqref{eq:bTE} both depend on the
$\mathcal{L}$ term, Eq.~\eqref{eq:Ldef}. For the sinusoidal profile,
the $\mathcal{L}$ term can be evaluated as
\begin{equation}
  \mathcal{L}_{m,m'}^{(+)} = 
  \imath^{m-m'} I_{m-m'}(\widetilde{\lambda}_{m} a).
\end{equation}
with $I_n$ the $n^{\text{th}}$ order modified Bessel functions of the
first kind.  Substituting $\vec{b}^{\text{TM}}_q$ and
$\vec{b}^{\text{TE}}_q$ in Eq.~\eqref{eq:Fmateq}, we construct an
$N\times N$ matrix of the $c$ coefficients with indices $m$ and
$q$. The $m$ index corresponds to the Fourier index, and runs from
$m=-M$ to $m=M$. The $q$ index corresponds to the index of the
eigenvalue and runs from $q=1$ to $q=N$.

It is now possible to obtain the Rayleigh coefficients from the $c$
ones by comparing Eq.~\eqref{eq:sol} and \eqref{eq:R_uw}. We can
identify the Rayleigh coefficients as proportional to the $c$
coefficients only if the eigenvalue $\lambda_q$ matches the Rayleigh
wavenumber $-\widetilde{\lambda}_m$. It should be noted, as mentioned
in Sec.~\ref{sec:perturb}, that the eigenvalue will perfectly match
the Rayleigh wavenumber only in the limit of inifinitely large
matrices. As is shown in Eq.~\eqref{eq:lam21}, only a subset of the
$q$ indexed eigenvalues will match with the Rayleigh wavenumbers for
finite matrices. Specifically the subset will match the Rayleigh
wavenumbers that correspond to Fourier modes with $m$ near zero. This
is clearly illustrated in the example given in
Fig.~\ref{fig:eigenvalues}. In this example, only the first 11
eigenvalues (circles) matched Rayleigh wavenumbers (crosses). In
practice, we consider that any eigenvalue matched a Rayleigh
wavenumber when relative difference between them was below some
tolerance ($10^{-3}$). We need to identify the $(q,m)$ pairs for which
$\lambda_q = \widetilde{\lambda}_m$. For the example given in
Fig.~\ref{fig:eigenvalues}, these are $(q,m)$ = (1,0), (2,-1), (3,1),
(4,-2), (5,2), etc. These matched $(q,m)$ pairs, along with the $c$
coefficients, the eigenvectors $\vec{V}$, and the $\mathcal{L}$ terms
are used in Eq.~\eqref{eq:Ray} to find the Rayleigh coefficients for
$m$ and $m'$. In the example, we would only keep $\mathbb{R}_{mm'}$
scattering coefficients for $m$ and $m'$ between $-5$ and $5$ as only
the first 11 eigenvalues matched with the Rayleigh wavenumbers.

In addition to Rayleigh coefficients
$\mathbb{R}$, we need to obtain $\mathbb{U}$ through Eq.~\eqref{eq:U} in
order to calculate the integrand given in Eq.~\eqref{eq:energy}. To obtain the Casimir energy, we then numerically integrate over wave-vectors
$\kappa$ and $k_x$ for a fixed maximum Fourier mode size $M$. All the
relevant matrices are calculated for $N=2M+1$ Fourier modes, running
from $m=-M$ to $M$. The maximum Fourier mode is steadily increased
from $M=1$ until the Casimir quantity of interest has converged. For
the purposes of this paper a Casimir quantity is defined as converged
when the relative change in the quantity when increasing $M$ by five
is less than $10^{-3}$.

\section{Results}
\label{sec:results}
\begin{figure}[htb]
  \begin{center}
  \includegraphics{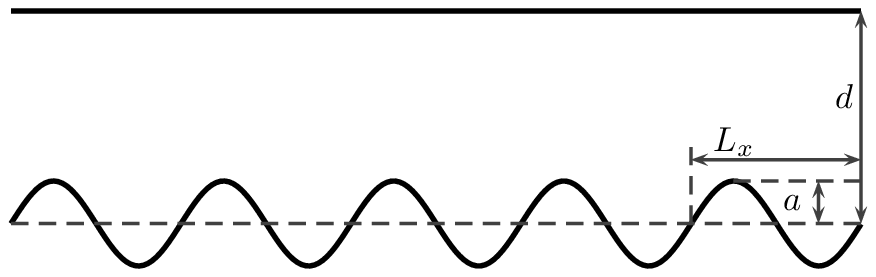}
  \caption{\label{fig:en} The test system for the energy
    calculations. The system consists of a flat plate separated from a
    sinusoidal grating with an average separation $d$. The sinusoidal
    grating has an amplitude $a$ and a wavelength $L_x$.}
  \includegraphics{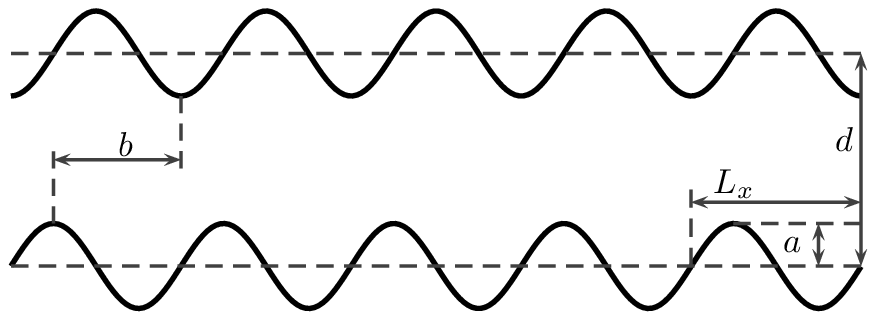}
  \caption{\label{fig:flat} The test system for the lateral force
    calculations. The system consists of two sinusoidal gratings, of
    equal amplitude $a$ and wavelength $L_x$. The gratings have an
    average separation $d$, and a lateral displacement between the
    peeks $b$.}
  \end{center}
\end{figure}
We now use the numerical method presented in the previous section to
calculate the energy and lateral force for two systems shown in
Figs.~\ref{fig:en} and \ref{fig:flat}. We calculate the Casimir energy
in the simplest system possible: a flat plate separated from a
sinusoidal grating (Fig.~\ref{fig:en}), and the lateral Casimir force
for the simplest system exhibiting a non-zero lateral force: two
identical sinusoidal gratings (Fig.~\ref{fig:flat}). The average
separation distance between two plates in both systems is $d$. The
sinusoidal gratings have an amplitude $a$ and a wavelength $L_x$ as
described in Figs.~\ref{fig:en} and \ref{fig:flat}.

\begin{figure}
  \begin{center}
  \includegraphics{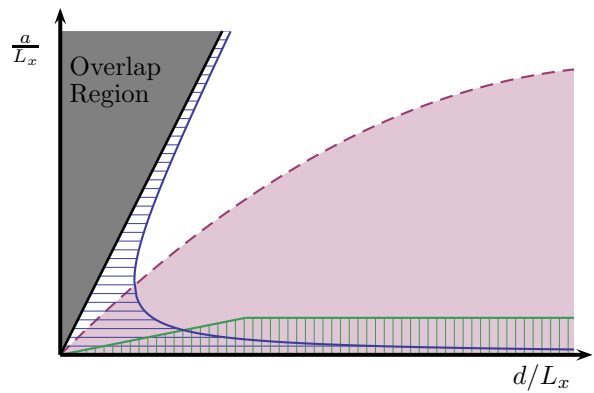}
  \caption{\label{fig:phasespace} (color online) Parameter space plot
    for the energy system shown in figure \ref{fig:en}. The two
    dimensionless quantities are the amplitude of corrugations divided
    by the wavelength $a/L_x$ and the separation divided by the
    wavelength $d/L_x$. The lighter (red) shaded region shows where
    for a fixed matrix size $N$, the C method gives converged results.
    The (blue) horizontally hashed region shows the region of validity
    for the PFA and DE. The (green) vertically hashed region shows the
    region of validity for the perturbative expansion.}
  \end{center}
\end{figure}

For the energy calculations, there are two dimensionless parameters,
$a/L_x$ the amplitude over the wavelength and $d/L_x$ the mean
separation over the wavelength.  The Casimir energy is calculated for
values of $a/L_x$ from 0 to $d/L_x$ (at which point the corrugations
would contact the planar surface) and for values of $d/L_x$ ranging
from $0.1$ to $5$. For the purpose of this paper, we increased the
value of the maximum Fourier mode $M$ from 1 up to 30 for the
convergence of the Casimir energy.  The Casimir energy converged for
$M \le 30$ for $a/L_x$ is less than 0.0575, 0.135, 0.4, 0.7, 1.2, 2.75
for values of $d/L_x=$ 0.1, 0.2, 0.5, 1, 2, 5 respectively.

For all tested values of $d/L_x$ the larger values of $a/L_x$ did not
converge for the maximum Fourier nodes $M<30$. It would be possible to
increase the range of convergence by choosing larger values of $M$, at
the cost of a longer computational time.  The lighter shaded (red
online) region in Fig.~\ref{fig:phasespace} illustrates the
approximate parameter space in which the Casimir calculation using the
C method converges.  The dashed (red) line is an empirical fit of the
data to a rightward facing parabola. The first obvious feature in the
figure is that if $a>d$ then the corrugations touch and pass through
the flat plate, an unphysical situation. This is shown in the
figure by the dark gray overlap region. The horizontal and vertical
hashed regions correspond to the regions of validity of the analytical
methods corresponding to the proximity force approximation with
derivative expansion and the perturbative expansion which we will
discuss in the next two sections, respectively.

A quick note should be stated about the computational expense of the
Casimir calculations using the C method. All numerics were programmed
in Wolfram Mathematica, on a modern desktop (2.8 GHz 64 bit processor,
with 8 GB ram). A single energy or lateral force calculation takes
about 100 cpu seconds for $M=10$ and about 900 cpu seconds (~15min)
for $M=30$. The computational cost appears to scale with $M^2$ for the
range of $M$ from 1 to 30.

\subsection{Comparison with PFA and Derivative Expansion}
\label{sec:pfacomp}

\begin{figure}
  \begin{center}
  \includegraphics[width=3.4in]{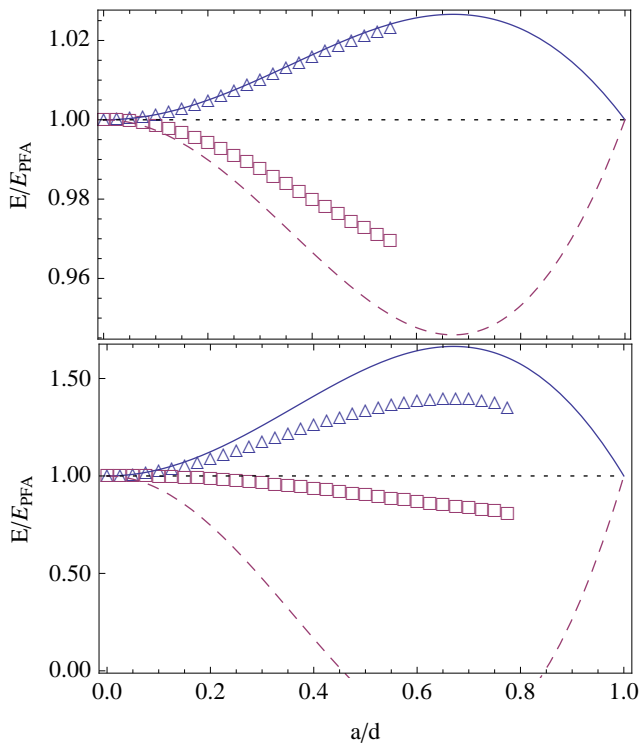}
  \caption{\label{fig:pfacomp}(color online) Casimir energy normalized
    to the PFA as defined by Eq.~\eqref{eq:pfa} versus $a/d$ for fixed
    $d/L_x=0.1$ (upper plot) and $d/L_x=0.5$ (lower plot). All
    parameters are defined in Fig.~\ref{fig:en}. The solid lines are
    the analytic formulas for the PFA plus the DE (Eqs.~\eqref{eq:pfa}
    and \eqref{eq:de}).  The solid (blue) curve corresponds to the TM
    mode, the dashed (red) curve corresponds to the TE mode. The
    triangles and squares correspond to the converged numerical
    results using the C method for the TM and TE modes respectively.}
  \end{center}
\end{figure}

The graphs in Fig.~\ref{fig:pfacomp} show the comparison of the
numerical calculations using the C method to the analytic results,
$E_{a}$ obtained as

\begin{equation}\label{eq:ea}
 E_a=E_{PFA}+E_{DE}
\end{equation}
with $E_{PFA}$ the PFA energy obtained by
assuming that the curved surfaces are made up of infinitesimal parallel
plates and summing over the contribution of all the plates. The
PFA approximation for the Casimir energy per unit area for a
sinusoidal grating as shown in Fig. \ref{fig:en} is
\begin{equation}\label{eq:pfa}
  \frac{E_{\text{PFA}}}{L_y L_x} = -\frac{\pi^2 \hbar c}{720} 
  \frac{2 d^2 + a^2}{2(d^2-a^2)^{5/2}}.
\end{equation}
The PFA is a valid approximation if the radius of curvature of objects
is large compared to the separation distance between the objects. For
our systems, this translates into the condition $(d-a) \ll L_x^2/a$,
given by the horizontally hashed region in Fig. \ref{fig:phasespace}.

The quantity $E_{DE}$ in Eq.~\eqref{eq:ea} corresponds to the
derivative expansion (DE) introduced in Ref.~\cite{Fosco:2011xx}
for scalar fields and Refs.~\cite{Bimonte:2012a,Bimonte:2012b}
for electric fields with perfect conductor or dielectric
boundaries. The first correction for the energy per unit area per mode
is
\begin{equation}\label{eq:de}
  \frac{E^p_{\text{DE}}}{L_y L_x} = 
  - \beta^p \frac{\pi^4 \hbar c}{360 L_x^2}
  \frac{a^2}{(d^2-a^2)^{3/2}},
\end{equation}
where $p$ indexes the polarization (TM or TE) and $\beta^p$ is a
constant given as $\beta^{\text{TM}}= 2/3$ and
$\beta^{\text{TE}}= 2/3 (1- 30/\pi^2)$.

The graphs in Fig.~\ref{fig:pfacomp} show the Casimir energy
normalized to the PFA vs $a/d$, with $a/d$ running
form 0 to 1 and for  two different values of $d/L_x$. The two curves in Fig.~\ref{fig:pfacomp} correspond a vertical trace in the parameter space in
Fig.~\ref{fig:phasespace}. The approximations should be exact at the
two limits of $a/d=0$ and $a/d=1$, related to a flat
plate and the tips of the corrugations touching the flat surface,
respectively. The upper graph is for the small fixed value $d/L_x=0.1$,
where it is expected that the PFA+DE approximations be fairly accurate
over the entire separation. The TM mode shows very good agreement over
the region of convergence. The DE for the TE mode seems to
overestimate the correction for this separation. The lower graph is
for a larger fixed value of $d/L_x=0.5$. For this separation the DE
seems to overestimate the correction to the PFA for both the TM and TE
modes over most of the range of convergence. It should be noted that
the numerical results for the TM mode show the beginning of a downward
turn as $a/d$ is increased.

\begin{figure}
  \begin{center}
  \includegraphics{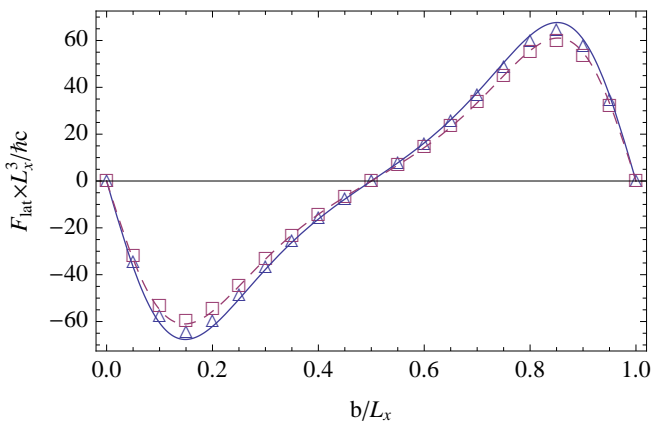}
  \caption{\label{fig:pfaforce}(color online) Lateral Casimir force
    per unit area in units of $\hbar c/ L_x^4$ vs lateral
    displacement. The curve is for a mean separation of $d/L_x=0.1$,
    and an amplitude of $a/d= 0.3$. The solid (blue) and dashed (red)
    lines correspond to the prediction of the PFA+DE for the TM and TE
    modes respectively. The triangles and squares correspond to the
    numerical results for the TM and TE modes, respectively.}
  \end{center}
\end{figure}
The graphs in Fig.~\ref{fig:pfaforce} show the lateral Casimir force
versus the lateral displacement for a fixed separation $d/L_x$ and
amplitude $a/d$. The values of the mean separation and amplitude were
chosen to show the range of applicability of the scattering technique
using the C method. The small value of the mean separation ensures the
validity of PFA+DE approximations, allowing a good benchmark to be
compared with our results. The larger relative value of $a/d$ should
ensure that the force is far from purely sinusoidal. The graph in
Fig.~\ref{fig:pfaforce} shows good agreement between the numerical
results and the analytic PFA+DE for both the TM and TE modes, and is
far from sinusoidal. For larger separations and larger amplitudes the
PFA+DE grossly overestimates the magnitude of the lateral Casimir
force.

\subsection{Comparison with Perturbative Calculations}
\label{sec:perturbcomp} 
This section compares the perturbative approximation derived in
section \ref{sec:perturb} to the numerical results. The perturbative
expansion results are obtained under the assumptions that $a/L_x \ll
1$ and $a/d \ll 1$. The region of validity for the perturbative
approximation is shown by the vertically hashed region in
Fig.~\ref{fig:phasespace}.

\begin{figure}
  \begin{center}
  \includegraphics{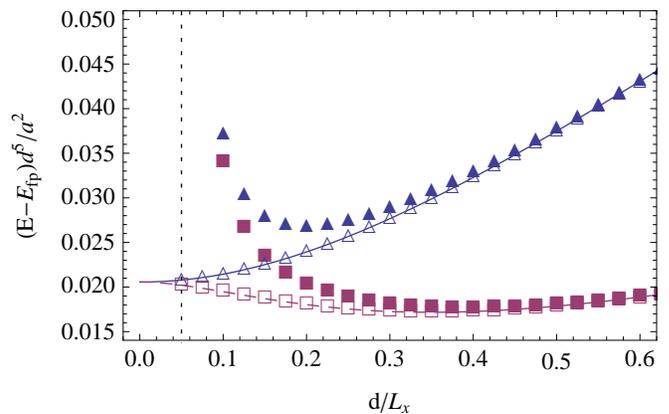}
  \caption{\label{fig:perturben} (color online) Correction to the
    Casimir energy in units of $\hbar c d^5/a^2$ versus $d/L_x$. The
    solid (blue) and dashed (red) lines are based on the analytical
    results given in Eq.~\eqref{eq:enp2} for the TM and TE modes,
    respectively. The empty triangles and squares are the numerical
    calculations using the C method for the TM and TE modes for fixed
    $a/d=0.1$, respectively. The filled triangles and squares are the
    numerical calculations using the C method for TM and TE modes for
    fixed $a/L_x=0.05$, respectively. The vertical dotted line is at
    $d/L_x=0.05$ and corresponds to the separation at which tips of
    the corrugations would touch the flat plate for $a/L_x=0.05$.}
  \end{center}
\end{figure}
A comparison of the perturbative expansion with the numerical
calculation using the C method is presented in
Fig.~\ref{fig:perturben}. The empty triangles are for a fixed small
value of $a/d=0.1$, which traces a diagonal line in
Fig.~\ref{fig:phasespace}. As shown in Fig.~\ref{fig:perturben}, for
this small value of $a/d$ the perturbative expansion should be valid
over the entire range of separations. The filled triangles are for a
fixed value of $a/L_x=0.05$, which traces a horizontal line in
Fig.~\ref{fig:phasespace}. As shown in Fig.~\ref{fig:perturben}, for
fixed profile heights the perturbative expansion breaks down and
severly underestimates the energy for short separations.  The
deviation of the numerical results from the perturbative expressions
becomes apparent for small values of $d/L_x$ as expected.

\begin{figure}
  \begin{center}
    \includegraphics{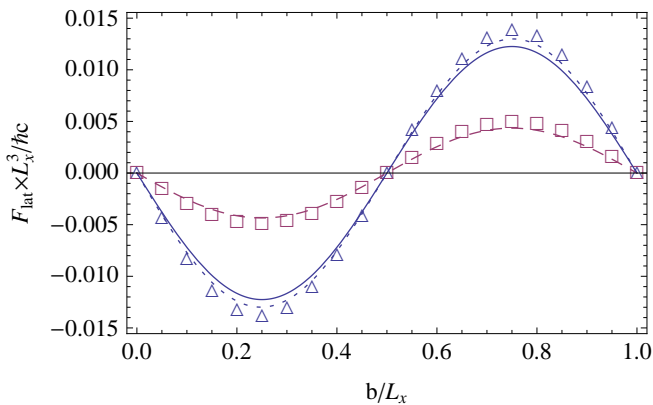}
    \caption{\label{fig:perturbforce} Lateral Casimir force per unit
      area in units of $\hbar c/L_x^4$ versus the lateral
      displacement. The triangles and squares are numerical
      calculations for TM and TE modes for constant mean separation
      $d/L_x=0.5$ and amplitude $a/d=0.1$. The solid lines are the
      second order perturbative expressions in Eq.~\eqref{eq:Flatteral} and the dashed line is the fourth order
      correction as calculated in Ref.~\cite{CaveroPelaez:2008tj} for
      the TM mode.}
  \end{center}
\end{figure}
The graph in Fig.~\ref{fig:perturbforce} illustrates a comparison of
the perturbative expansion results with the numerical calculation for
the lateral Casimir force. The calculation is done for a larger value
of mean separation $d/L_x=0.5$ and a smaller value of the amplitude
$a/L_x=0.05$. In this region of parameter space one expects the
perturbative expansion to be still valid, while the PFA begins to
fail. The graph shows good agreement with the perturbative
approximation for both TM (triangles) and TE (squares) modes. The
dashed line represents the next to leading order expansion as
calculated in Refs.~\cite{CaveroPelaez:2008tj}. For larger amplitudes,
the perturbative expansion dramatically underestimates the lateral
Casimir force, not shown in the figure.

\section{Conclusion}\label{sec:conclusion}
In general, it is difficult to calculate the Casimir energy between
non-planar surfaces, and the analytical approximations used such as
PFA, the derivative expansion, and perturbative expansions in the
profile height are all valid in a small region of the parameter space,
see Fig.~\ref{fig:phasespace}. In this paper, we combined the
scattering theory with the C method, a powerful technique borrowed
from electromagnetic grating theory, to calculate the Casimir energy
for surface relief gratings.

The C method is in particular suitable for calculating the scattering
matrices for smoothly varying height profiles, which allows us to
obtain the Casimir energy over a wider range of phase space than that
accessible by the analytic approximations.  Figure
\ref{fig:phasespace} shows the region of convergence of C method when
we kept the size of our relevant matrices relatively small. Each point
obtained in the plots of Figs.~\ref{fig:pfacomp}, \ref{fig:pfaforce},
\ref{fig:perturben}, and \ref{fig:perturbforce} was produced using a
desktop computer for ~900 cpu seconds. One can easily expand the
region of convergence by increasing the size of matrices and
allocating more CPU time. The limiting factor will then become the
roundoff error inherent to the solution of the system of equations
given in \eqref{eq:Fmateq}. This issue has been addressed in
Ref.~\cite{Poyedinchuk:2006a} along with an alternative method for
solving the system of equations that improves numerical stability.

It is important to note that the scattering method has been previously
used to calculate the Casimir interactions for surface relief gratings
using a small amplitude perturbation series for the reflection
coefficients\cite{Neto:2005zz, Rodrigues:2006ku, Rodrigues:2006cf}.
We emphasize that in all previous work, in order to find the
perturbative corrections to the reflection coefficients the Rayleigh
hypothesis was assumed.  In addition to the scattering approach, other
techniques were employed to calculate the Casimir forces for perfect
materials\cite{Emig:2001dx, Emig:2002xz, CaveroPelaez:2008tj,
  CaveroPelaez:2008rt}. These calculations were also done under the
assumption of the Rayleigh hypothesis. In this paper, we performed all
the calculations without assuming the Rayleigh hypothesis and showed
that the perturbative calculations implicitly make such an assumption.

We compared the numerical results of this paper against known analytic
approximations: the PFA plus the first correction to the PFA using the
derivative expansion and the perturbative expansion and found very
good agreement in the regions of parameter space where the
approximations are expected to be valid.

The method presented in the paper can be easily extended beyond the
perfectly conducting case to include general dielectrics, and is
valuable tool to employ in understanding the Casimir force for surface
relief gratings.

\section*{Acknowledgments}
The authors would like to thank Alex Banishev, Umar Mohideen, Ehsan
Noruzifar, Thorsten Emig, and Mehran Kardar, Kim Milton, Prachi
Parashar, and Elom Abalo for helpful discussions. JW thanks Stephen
Fulling for the invitation to present this work at Texas A\&M. This
work was supported by DMR-1310687.

\appendix

\section{Perturbative Calculation}\label{app:1}
The appendix provides some of the details of the derivations of the
perturbative results found in the Sec.~\ref{sec:perturb}. The goal is
to find an analytic approximation for the Casimir energy or lateral
Casimir force for surface relief gratings. The Casimir energy and
lateral Casimir force depend upon the Rayleigh coefficients, which in
turn depend upon the eigenvectors and eigenvalues of the quadratic
eigenvalue problem given in Eq.~\eqref{eq:qep}. This section will
cover the results in the order they are needed in the body of the
paper: First we expand the eigenvectors and eigenvalues, second the Rayleigh
coefficients, and finally the Casimir energy and lateral Casimir
force up to the second order in the amplitude of height profile.

\subsection{Eigenvalue and Eigenvectors} \label{app:ev}
The quadratic eigenvalue problem in Eq.~\eqref{eq:qep} can be solved
perturbatively by expanding the elements of $\mat{A}_0$ and
$\mat{A}_1$ matrices and the eigenvalues $\lambda_q$ and eigenvectors
$\vec{V}_q$ in powers of the profile functions $h(x)$. It should be
noted that in the perturbative regime, it is possible to immediately
identify the $q$ index with a Fourier mode, so for the rest of the
appendix we will replace all $q$ indices with $m$ ones. We have
already presented the zeroth order quadratic eigenvalue equation
\eqref{eq:zero}, in Sec.~\ref{sec:perturb}.  The first order equation
can be easily found to be
\begin{multline}\label{eq:evone}
  -\big(\big(\lambda^{(0)}_m\big)^2
  - \mat{A}_0\big) \cdot \vec{V}^{(1)}_m \\
  - \big(2 \lambda^{(0)}_m\lambda^{(1)}_m + 
  \lambda^{(0)}_m \mat{A}_1\big) \cdot \vec{V}^{(0)}_m = 0.
\end{multline}
If we multiply Eq.~\eqref{eq:evone} by $\vec{V}^{(0)}_{m}$ ($m'=m$)
then the first term will give exactly zero, as shown in
Eq.~\eqref{eq:zero}.  We can find the first correction for the
eigenvalue
\begin{equation}
  \lambda^{(1)}_m = \frac{1}{2}\big(\mat{A}_1\big)_{mm}
\end{equation}
Note that the $m,m'$ elements of the $\mat{A}_1$ matrix, given explicity in
Eqs.~\eqref{eq:gh} and \eqref{eq:A1mat} , are in turn proportional to
the inverse lattice vector $\vec{G}_{m-m'}$. Using Eq.~\eqref{eq:G},
we can show that the diagonal elements of the $\mat{A}_1$ matrix are
zero. Thus, the first correction to the eigenvalue is zero as given in
Eq.~\eqref{eq:lam1}.

By multiplying Eq.~\eqref{eq:evone} with $\vec{V}^{(0)}_{m'}$ for $m
\ne m'$ and using the identifcation that the zeroth order eigenvalues
are Kronecker delta functions, we can idenify the $m \ne m'$
components of the first correction to the eigenvector. The off
diagonal components of $\big(\vec{V}^{(1)}_m\big)_m'$ are proportional
to the off diagonal elements of $\mat{A}_1$ and to the difference in the
Rayleigh wavenumbers
\begin{equation}\label{eq:v1partial}
  \big(\vec{V}^{(1)}_m\big)_{m'} = \frac{-\widetilde{\lambda}_m
    \big(\mat{A}_1\big)_{m'm} } { \widetilde{\lambda}_{m'}^2
    -\widetilde{\lambda}_m^2}.
\end{equation}
Inserting Eqs.~\eqref{eq:Kmat} and \eqref{eq:Ghmat} into Eq.~\eqref{eq:A1mat} for the matrix elements of $\big(\mat{A}_1\big)$ and
using the definition of the Rayleigh eigenvalues from Eq.~\eqref{eq:ltdef}, we find
\begin{subequations}\label{eq:v1ident}
\begin{align}
  \big(\mat{A}_1\big)_{m'm} & =
  \vec{K}_{m'+m}\vec{G}_{m'-m}h_{m'-m},\\
  \widetilde{\lambda}_{m'}^2 - 
  \widetilde{\lambda}_m^2 &
  = \vec{K}_{m'+m}\vec{G}_{m'-m},
\end{align}
\end{subequations}
Substituting Eqs.~\eqref{eq:v1ident} in
Eq.~\eqref{eq:v1partial}, a simplified version of the first order
correction to the eigenvectors can be found, and is given in
Eq.~\eqref{eq:V1}.

The second order equation in the pertubative expansion of the
quadratic eigenvalue problem in Eq.~\eqref{eq:qep} is
\begin{multline}
  -\big(\big(\lambda^{(0)}_m\big)^2
  - \mat{A}_0\big) \cdot \vec{V}^{(2)}_m
  - \lambda^{(0)}_m \mat{A}_1 \cdot \vec{V}^{(1)}_m \\
  + \big( \big(\lambda^{(0)}_m\big)^2 \mat{A}_2
  -2 \lambda^{(0)}_m\lambda^{(2)}_m\big) \cdot \vec{V}^{(0)}_m= 0.
\end{multline}
Proceeding in the same manor as for the first order equation, we find
the second order corrections for the eigenvalues and eigenvectors given
in Eqs.~\eqref{eq:lam21} and \eqref{eq:V2} respectively.

\subsection{Reflection Coefficients}\label{app:refl}
The next task is to find the perturbative expression for the Rayleigh
coefficients. We have shown in Eqs.~\eqref{eq:Ray} and \eqref{eq:VeqL} that the Rayleigh coefficients are equivalent to the
undetermined $c$ constants in the general solutions found using the
C-method, Eq.~\eqref{eq:sol}. It is then possible to perturbatively solve the
Rayleigh coefficients through
Eq.~\eqref{eq:Fmateq}. In the same manor as with the eigenvalues and
eigenvectors, we expand the Rayleigh coefficients in a series,
\begin{equation}\label{eq:Rexp}
  \mathbb{R}^{p}_{mm'} = \sum_{i} \mathbb{R}^{p(i)}_{mm'},
\end{equation}
where the $(i)$ index implies the term is of order
$\mathcal{O}(h^i)$. Substituting equations \eqref{eq:Rexp} into
Eq.~\eqref{eq:Fmateq} and expanding $\mat{F}$ matrix and $\vec{b}$, we
gather terms by powers of the height profile function $h$. The resulting set of systems of equations can then be solved iteratively to find
the Rayleigh coefficients for both the $TM$ and $TE$ polarizations.

The TM and TE modes obey different systems of equations. The system of
equations for the zeroth order TM mode is
\begin{equation}\label{eq:Rtmzeroeq}
  \sum_{m''} \big(\vec{V}^{(0)}_{m''}\big)_{m'}
  \mathbb{R}^{TM(0)}_{mm''} 
  = - \mathcal{L}^{(+)(0)}_{mm'},
\end{equation}
which can be solved to find the result given in Eq.~\eqref{eq:Rtm0}.  Next we find the
first order system of equations for the TM mode as
\begin{equation}\label{eq:Rtmoneeq}
  \sum_{m''}
  \Big[ \big(\vec{V}^{(0)}_{m''}\big)_{m'} 
    \mathbb{R}^{TM(1)}_{mm''} +
    \big(\vec{V}^{(1)}_{m''}\big)_{m'} \mathbb{R}^{TM(0)}_{mm''} \Big]=
    - \mathcal{L}^{(+)(1)}_{mm'}.
\end{equation}
Inserting Eqs.~\eqref{eq:V1}, \eqref{eq:Lexp} and \eqref{eq:Rtm0}
into Eq.~\eqref{eq:Rtmoneeq}, and solving for $\mathbb{R}^{TM(1)}$
we find Eq.~\eqref{eq:Rtm1}.  

The second order system of equations for
the TM is
\begin{multline}\label{eq:Rtmtwoeq}
  \sum_{m''}\Big[ \big(\vec{V}^{(0)}_{m''}\big)_{m'}
    \mathbb{R}^{TM(2)}_{mm''} + \big(\vec{V}^{(1)}_{m''}\big)_{m'}
    \mathbb{R}^{TM(1)}_{mm''} \\ + \big(\vec{V}^{(2)}_{m''}\big)_{m'}
    \mathbb{R}^{TM(0)}_{mm''} \Big] = - \mathcal{L}^{(+)(2)}_{mm'}.
\end{multline}
Using Eqs.~ \eqref{eq:Lexp},\eqref{eq:VeqL} and \eqref{eq:Rtm0}, we
find the expressions for $\vec{V}^{(2)}\mathbb{R}^{TM(0)}$ and
$\mathcal{L}^{(+)(2)}$ are identical with the same sign, so they
cancel exactly. The expression for $\mathbb{R}^{TM(2)}$ only depends
upon the second term in Eq.~\eqref{eq:Rtmtwoeq} and is explicitly
given in Eq.~\eqref{eq:Rtm2}.

The zeroth order system of equation for the TE mode is
\begin{equation}\label{eq:Rtezeroeq}
  \sum_{m''} (-\widetilde{\lambda}_{m''})
  \big(\vec{V}^{(0)}_{m''}\big)_{m'} \mathbb{R}^{TE(0)}_{mm''} 
  = - \widetilde{\lambda}_m \mathcal{L}^{(+)(0)}_{mm'}.
\end{equation}
which can be solved to find $\mathbb{R}^{TE(0)}$ given explicitly in
Eq.~\eqref{eq:Rte0}. The first order system of equation for the TE
mode is
\begin{multline}\label{eq:Rteoneeq}
  \sum_{m''} \Big[ \big(\mat{Gh}\cdot\mat{K}\cdot\vec{V}^{(0)}_{m''}
    -\widetilde{\lambda}_{m''}\vec{V}^{(1)}_{m''}\big)_{m'}
    \mathbb{R}^{TE(0)}_{mm''}
    \\ -\widetilde{\lambda}_{m''}\big(\vec{V}^{(0)}_{m''}\big)_{m'}
    \mathbb{R}^{TE(1)}_{mm''}\Big] = - \widetilde{\lambda}_m
  \mathcal{L}^{(+)(1)}_{mm'}\\ - \sum_{m''} \big( \mat{Gh}\cdot
  \mat{K}\big)_{m''m'} \mathcal{L}^{(+)(0)}_{mm''}.
\end{multline}
Substituting Eqs.~\eqref{eq:gh}, \eqref{eq:Lexp}, \eqref{eq:VeqL} and
\eqref{eq:Rte0} into Eq.~\eqref{eq:Rteoneeq} and solving for
$\mathbb{R}^{TE(1)}$ we find the expression for the Rayleigh
coefficient \eqref{eq:Rte1}.  The second order system of equations for
the TE mode contains 9 terms overall, however it can be very quickly
simplified. In a cancellation similar to what occurred for the TM
case, three terms on the left hand side exactly cancel with the three
terms on the right hand side. The simplified system of equations is
\begin{multline}\label{eq:Rtetwoeq}
  \sum_{m''} \Big[
    \big(\mat{Gh}\cdot\mat{K}\cdot\vec{V}^{(0)}_{m''}
    -\widetilde{\lambda}_{m''}\vec{V}^{(1)}_{m''} \big) 
    \mathbb{R}^{TE(1)}_{mm''}\\
    -\widetilde{\lambda}_{m''}\vec{V}^{(0)}_{m''} 
    \mathbb{R}^{TE(2)}_{mm''}\Big] = 0.
\end{multline}
This is solved to give equation \eqref{eq:Rte2}.

\subsection{Casimir energy and lateral Casimir force}
\label{app:en}
This section describes in more detail the derivation of the
perturbative expansions of the Casimir energy and lateral force.  In
the scattering method the Casimir energy is  proportional to
an expression of the form
\begin{equation}\label{eq:EMmat}
  E \propto \Tr \ln \big(1-\mat{M}\big),
\end{equation}
with $\mat{M}$ a matrix as given in Eq.~1. Let the matrix $M$ be
slightly perturbed such that $\mat{M} \to \mat{M}+ \mat{dM}$. Equation
\eqref{eq:EMmat} then yields
\begin{equation}\label{eq:enexp1}
  E \propto \Tr \ln \big(1-\mat{M}) + \Tr \ln
  \bigg(1-\frac{\mat{dM}}{1-\mat{M}}\bigg),
\end{equation}
The second term can be considered as a perturbation to the energy
$dE$. Because the $\mat{dM}$ term is a small perturbation (compared to
$\mat{M}$) the logarithm in equation \eqref{eq:enexp1} can be expanded
to yield
\begin{equation}
  dE \propto - \sum_s \frac{1}{s} 
  \Tr \bigg(\frac{\mat{dM}}{1-\mat{M}}\bigg)^s.
\end{equation}
The individual terms in the series can easily be thought of as the
coefficients of a perturbation series in $dM$.

To calculate the energy, we consider the system shown in
Fig.~\ref{fig:en}. Since one of the surfaces is a flat plate, we need
to find the Rayleigh coefficients for a flat plate. The plane wave
reflection coefficients for perfectly conducting flat plate are known
to be $r^{TM}=-1$, and $r^{TE}=-1$. By switching from the plane wave
basis to a Block wave basis given in Eq.~\eqref{eq:fscatprob} we can
identify the Rayleigh coefficients for flat plates as
$\mathbb{R}^{TM}_{mm'}=-\delta_{mm'}$ and
$\mathbb{R}^{TE}_{mm'}=\delta_{mm'}$. The Casimir energy from
Eq.~\eqref{eq:en1} can be rewritten as
\begin{equation}\label{eq:cas}
  \frac{E}{L_y L_x} = \frac{\hbar c}{8 \pi^2}
  \int_0^\infty \!\!\!\!\!\! \kappa \dif \kappa \!\!
  \int_{-\pi/L_x}^{\pi/L_x} \!\!\!\!\!\!\!\! \dif k_x 
  \sum_p \ln\det\big(1\pm\mathbb{R}^{p}|\mathbb{U}^{12}|^2\big),
\end{equation}
where the $+$ or $-$ is for the TM or TE mode respectively.

For a flat plate and single periodic grating
$\mat{M}=\mathbb{R}|\mathbb{U}|^2$. The perturbation to the full
matrix is written in terms of perturbation to the Rayleigh coefficients
$\mat{dM} = d\mathbb{R} |\mathbb{U}|^2$, where the corrections up to
the second order are included $d\mathbb{R} = \mathbb{R}^{(1)} +
\mathbb{R}^{(2)}$. Using this formula the first order correction to
the energy is
\begin{multline}
  \frac{E^{(1)}}{L_y L_x} =\\ -\frac{\hbar c}{8 \pi^2} 
  \int_0^\infty \!\!\!\!\! \kappa \dif \kappa
  \int_{-\pi/L_x}^{\pi/L_x} \!\!\!\!\!\!\!\! \dif k_x 
  \Tr \bigg(\frac{\mathbb{R}^{(1)}|\mathbb{U}|^2}{
    1-\mathbb{R}^{(0)}|\mathbb{U}|^2} \bigg).
\end{multline}
Using Eqs.~\eqref{eq:Rtmexp} and \eqref{eq:Rteexp} for the Rayleigh
coefficients and considering the translation matrix
\begin{equation}\label{eq:Ufmat}
  \mathbb{U}_{mm'} = \delta_{mm'} 
  e^{\imath \vec{K}_m b - \widetilde{\lambda}_m d},
\end{equation}
the first order correction to the energy becomes
\begin{multline} \label{eq:e1}
  \frac{E^{(1)}}{L_y L_x} = \\
-\frac{\hbar c}{8 \pi^2} 
  \int_0^\infty \!\!\!\!\! \kappa \dif \kappa
  \int_{-\pi/L_x}^{\pi/L_x} \!\!\!\!\!\!\!\! \dif k_x 
  \sum_m \frac{2 h_0 \widetilde{\lambda}_m 
    e^{-2 \widetilde{\lambda}_m d}}{
    1-e^{-2 \widetilde{\lambda}_m d}}.
\end{multline}
Since $\widetilde{\lambda}_m$ is a function of
$k_x+ 2\pi m/L_x$, it is possible to simplify Eq.~\ref{eq:e1}
using the identity
\begin{equation}
  \int_{-a}^a \!\!\!\! \dif x \sum_m f(x+2 m a) = 
  \int_{-\infty}^\infty \!\!\!\!\!\! \dif x \; f(x),
\end{equation}
which can be even further simplified by changing variables
to polar coordinates
\begin{subequations}
\begin{align}
  \kappa &= \lambda \cos \beta, &
  k_x &= \lambda \sin \beta.
\end{align}
\end{subequations}
Considering $\kappa^2+k_x^2 = \widetilde{\lambda}_0^2 = \lambda^2$ the
first correction to the energy can be written
\begin{equation}
  \frac{E^{(1)}}{L_y L_x} = -\frac{\hbar c h_0}{4 \pi^2} 
  \int_{-\pi/2}^{\pi/2} \!\!\!\!\!\!\!\! \cos \beta \dif \beta 
  \int_0^\infty \!\!\!\!\! \lambda^2 \dif \lambda
  \frac{\lambda e^{-2 \lambda d}}{1-e^{-2\lambda d}}.
\end{equation}
The integrals can be evaluated exactly to yield the expression
obtained in Eq.~\eqref{eq:enp1} in section \ref{sec:perturb}.

The second order correction to the energy is written in terms of the
perturbative Rayleigh coefficients as
\begin{multline}
  \frac{E^{(2)}}{L_y L_x} = -\frac{\hbar c}{8 \pi^2} 
  \int_0^\infty \!\!\!\!\! \kappa \dif \kappa
  \int_{-\pi/L_x}^{\pi/L_x} \!\!\!\!\!\!\!\! \dif k_x\\ 
  \bigg[ \Tr \bigg(\frac{\mathbb{R}^{(2)}|\mathbb{U}|^2}{
      1-\mathbb{R}^{(0)}|\mathbb{U}|^2}\bigg) + \frac{1}{2}
    \Tr \bigg(\frac{\mathbb{R}^{(1)}|\mathbb{U}|^2}{
      1-\mathbb{R}^{(0)}|\mathbb{U}|^2}\bigg)^2 \bigg].
\end{multline}
Upon substituting Eqs.~\eqref{eq:Rtmexp} and \eqref{eq:Rteexp} for the
Rayleigh coefficients and using the translation matrix in
Eq.~\eqref{eq:Ufmat}, we find
\begin{multline}
  \frac{E^{(2)\text{TM}}}{L_y L_x} = -\frac{\hbar c}{4 \pi^2} 
  \int_0^\infty \!\!\!\!\! \kappa \dif \kappa
  \int_{-\pi/L_x}^{\pi/L_x} \!\!\!\!\!\!\!\! \dif k_x\\ 
  \sum_{m,m'} |h_{m-m'}|^2 \frac{\widetilde{\lambda}_m 
    e^{-2 \widetilde{\lambda}_m d}}{
    1-e^{-2 \widetilde{\lambda}_m d}}
  \bigg(\frac{ \widetilde{\lambda}_{m'}}{1-
    e^{-2 \widetilde{\lambda}_{m'} d}}\bigg),
\end{multline}
and
\begin{multline}
  \frac{E^{(2)\text{TE}}}{L_y L_x} = -\frac{\hbar c}{4 \pi^2} 
  \int_0^\infty \!\!\!\!\! \kappa \dif \kappa
  \int_{-\pi/L_x}^{\pi/L_x} \!\!\!\!\!\!\!\! \dif k_x\\ 
  \sum_{m,m'} |h_{m-m'}|^2 \frac{ 
    e^{-2 \widetilde{\lambda}_m d}}{
    1-e^{-2 \widetilde{\lambda}_m d}}
  \bigg(\frac{ \widetilde{\lambda}_{mm'}^2}{1-
    e^{-2 \widetilde{\lambda}_{m'} d}}\bigg).
\end{multline}
These expressions can also be simplified using the identity
\begin{multline}\label{eq:ident2}
  \int_{-a}^a \!\!\!\!\! \dif x 
  \sum_m \sum_{m'} B_{m-m'} f(x + 2 m a, x + 2 m' a) = \\
  \sum_{m'} B_{m-m'} \int_{-\infty}^\infty \!\!\!\!\! \dif x \;
  f(x, x + 2 (m'-m) a).
\end{multline}
Upon a change of variable to the polar coordinates and setting $z=2\lambda d$, the
second order contribution for the energy can be written as given in Eq.~\eqref{eq:enp2} with $g_p$ terms given as
\begin{subequations}\label{eq:gterms}
\begin{equation}
  g_{\text{TM}}(A) = \frac{15}{8\pi^4}
  \int_0^\infty \!\!\!\!\! \dif z 
  \frac{z^2 e^{-z}}{1-e^{-z}}
  \int_{-1}^1 \!\!\!\! \dif x 
  \frac{z'}{1-e^{-z'}},
\end{equation}
and
\begin{equation}
  g_{\text{TE}}(A) = \frac{15}{8\pi^4}
  \int_0^\infty \!\!\!\!\! \dif z 
  \frac{z^2 e^{-z}}{1-e^{-z}}
  \int_{-1}^1 \!\!\!\! \dif x 
  \frac{(z + A x)^2}{z'(1-e^{-z'})},
\end{equation}
\end{subequations}
with $z'= \sqrt{z^2 + A^2 + 2 z A x}$.

The prefactor $15/8\pi^4$ is chosen such that the functions are
normalized for zero argument $g_p(0)=1$. In the limit of large
argument the function have linear asymptotic behavior
$g_{TM}(A) \sim \frac{A}{4}$, and $g_{TE}(A) \sim \frac{A}{12}$.

In order to find the lateral force we will examine the system between
two surface relief gratings labeled by superscripts 1 and 2 separated
by a distance $d$ with a distance $b$ between peaks as shown in
Fig.~\ref{fig:flat}. In order to find the lateral force, we will first
find the energy, and take the derivative with respect to $b$. For two
corrugated surfaces the matrix in Eq.~\eqref{eq:EMmat} is $\mat{M}=
\mathbb{R}^1 \mathbb{U}^{12} \mathbb{R}^2 \mathbb{U}^{21}$. The
perturbed matrix would be then written
\begin{multline}
  \mat{M} = d\mathbb{R}^1\mathbb{U}
  \mathbb{R}^{(0)}\mathbb{U}^\dagger + \mathbb{R}^{(0)}\mathbb{U}
  d\mathbb{R}^2\mathbb{U}^\dagger + \\ d\mathbb{R}^1\mathbb{U}
  d\mathbb{R}^2\mathbb{U}^\dagger,
\end{multline}
where the perturbation to the Rayleigh coefficients contains the first
two terms of the perturbative expansion $d\mathbb{R}^i =
\mathbb{R}^{i(1)}+ \mathbb{R}^{i(2)}$ and the translation matrix from
surface 2 to 1 has been identified as the conjugate of the translation
matrix from 1 to 2. The surface index $i$ has been dropped from the
$R^{(0)}$ term because in zeroth order both surfaces are describe by
flat plates. The perturbed energy can contain contributions from
either the first surface, the second surface, or both. Only terms that
contain contributions from both surfaces will contribute to the
lateral force. The lowest order mixed term in the energy is 
\begin{multline}
  \frac{E^{(2)}}{L_y L_x} = - \frac{\hbar c}{8\pi^2}
  \int_0^\infty \!\!\!\!\! \kappa \dif \kappa
  \int_{-\pi/L_x}^{\pi/L_x} \!\!\!\!\!\!\!\! \dif k_x \\ 
  \bigg[ \Tr \bigg(\frac{\mathbb{R}^{1(1)}\mathbb{U}
      \mathbb{R}^{2(1)}\mathbb{U}^\dagger}{
      1-\mathbb{R}^{(0)}\mathbb{U}\mathbb{R}^{(0)}
      \mathbb{U}^\dagger}\bigg) + \\
  \Tr \bigg(\frac{\mathbb{R}^{1(1)}\mathbb{U}|\mathbb{U}|^2 
    \mathbb{R}^{2(1)}\mathbb{U}^\dagger}{
      (1-\mathbb{R}^{(0)}\mathbb{U}
    \mathbb{R}^{(0)}\mathbb{U}^\dagger})^2\bigg)^2 \bigg].
\end{multline}
Substituting Eqs.~\eqref{eq:Rtmexp} and \eqref{eq:Rteexp} into
Eq.~\eqref{eq:Ufmat}, we find the correction to the energy for the TM and TE
modes as
\begin{subequations}
\begin{multline}
  \frac{E^{(2)\text{TM}}}{L_y L_x} = - \frac{2 \hbar c}{\pi^2}
  \int_0^\infty \!\!\!\!\! \kappa \dif \kappa
  \int_{-\pi/L_x}^{\pi/L_x} \!\!\!\!\!\!\!\! \dif k_x \\ 
  \sum_{mm'} h^1_{m'-m}h^2_{m-m'} e^{\imath G_{m'-m} b} \\
  \frac{\widetilde{\lambda}_m \widetilde{\lambda}_{m'}}{
    \sinh (\widetilde{\lambda}_m d)\sinh(\widetilde{\lambda}_{m'} d)},
\end{multline}
and
\begin{multline}
  \frac{E^{(2)\text{TE}}}{L_y L_x} = - \frac{2 \hbar c}{\pi^2}
  \int_0^\infty \!\!\!\!\! \kappa \dif \kappa
  \int_{-\pi/L_x}^{\pi/L_x} \!\!\!\!\!\!\!\! \dif k_x \\ 
  \sum_{mm'} h^1_{m'-m}h^2_{m-m'} e^{\imath G_{m'-m} b} \\
  \frac{\widetilde{\lambda}^4_{mm'}/
    \widetilde{\lambda}_m \widetilde{\lambda}_{m'}}{
    \sinh (\widetilde{\lambda}_m d)\sinh(\widetilde{\lambda}_{m'} d)}.
\end{multline}
\end{subequations}
These expressions can be further simplified using the identity
\eqref{eq:ident2}. After
taking the derivative with respect to $b$ the lateral force becomes
\begin{multline}\label{eq:Flatteral}
  \frac{F^{(2)p}}{L_y L_x} = - \frac{\hbar c \pi^3}{240} 
  \frac{1}{d^3 L_x} \sum_{m=1}^\infty j_p\big(\tfrac{2\pi m d}{L_x}\big) \\
  \Big(\operatorname{Re}\big(h^1_m h^2_{-m}\big) 
  \sin \! \big(\tfrac{2\pi m b}{L_x}\big) \\
  \!+ \! \operatorname{Im}\big(h^1_m h^2_{-m}\big) 
  \cos \! \big(\tfrac{2\pi m b}{L_x}\big) \Big).
\end{multline}
The $j_p$ functions are 
\begin{subequations}
\begin{equation}
j_{\text{TM}}(A) = \frac{60}{\pi^4}
\int_0^\infty \!\!\!\!\! \dif z \; \frac{z^3}{\sinh z}
\int_{-1}^1 \!\!\!\! \dif x \;\frac{z'}{\sinh z'},
\end{equation}
and
\begin{equation}
j_{\text{TE}}(A) = \frac{60}{\pi^4}
\int_0^\infty \!\!\!\!\! \dif z \; \frac{z}{\sinh z}
\int_{-1}^1 \!\!\!\! \dif x \;\frac{(z^2 +z A x)^2}{z' \sinh z'},
\end{equation}
\end{subequations}
with $z'= \sqrt{z^2 + A^2 + 2 z A x}$. After a proper scaling, the $j_p$ functions match the $J_p$ functions in equations (46), (47), and (48) in
Ref.~\cite{Emig:2002xz}. Also, the
integral expression for $j_{\text{TM}}$ for the TM mode can be shown
to be identical to the $A^{(1,1)}_D(x)$ expression in equation (61) in
Ref.~\cite{CaveroPelaez:2008tj}.

\bibliography{cmethod.bib}

\end{document}